\documentclass[useAMS,usenatbib]{mn2e}

\newif\ifAMStwofonts

\usepackage{graphicx}
\usepackage[usenames,dvipsnames]{color}

\title[X-ray polarimetry \& the Galactic Centre]
      {Prospects of 3D mapping of the Galactic Centre clouds with X-ray polarimetry}
\author[Marin et al.]
      {F.~Marin$^1$\thanks{E-mail: frederic.marin@asu.cas.cz},
       V. Karas$^1$, D. Kunneriath$^1$ and F. Muleri$^2$\\
       $^1$Astronomical Institute of the Academy of Sciences, Bo{\v c}n\'{\i} II 1401, CZ-14100 Prague, Czech Republic\\
       $^2$INAF/IAPS, Via del Fosso del Cavaliere 100, I-00133 Roma, Italy}

\date{Accepted 2014 April 10.  Received 2014 March 27; in original form 2014 February 20}

\pagerange{\pageref{firstpage}--\pageref{lastpage}}
\pubyear{2014}

\begin{document}

\maketitle

\label{firstpage}

\begin{abstract}
Despite past panchromatic observations of the innermost part of the Milky Way, the overall structure of the Galactic Centre (GC)
remains enigmatic in terms of geometry. In this paper, we aim to show how polarimetry can probe the three-dimensional position of the 
molecular material in the central $\sim$100~pc of the GC. We investigate a model where the central supermassive black hole Sgr~A$^*$ is 
radiatively coupled to a fragmented circumnuclear disc (CND), an elliptical twisted ring representative of the central molecular zone 
(CMZ), and the two main, bright molecular clouds Sgr~B2 and Sgr~C. 8 -- 35~keV integrated polarization mapping reveals that Sgr~B2 
and Sgr~C, situated at the two sides of the CMZ, present the highest polarization degrees (66.5 and 47.8 per cent respectively), both 
associated with a polarization position angle $\psi$~=~90$^\circ$ (normal to the scattering plane). The CND shows a lower polarization 
degree, 1.0 per cent with $\psi$~=~-20.5$^\circ$, tracing the inclination of the CND with respect to the Galactic plane. The CMZ polarization 
is spatially variable. We also consider a range of spatial locations for Sgr~A$^*$ and the reprocessing media, and investigate how the 
modeled three-dimensional geometry influences the resulting GC polarization. The two reflection nebulae are found to always produce high 
polarization degrees ($\gg$~10 per cent). We show that a 500~ks observation with a broadband polarimeter could constrain the location and 
the morphology of the scattering material with respect to the emitting source, revealing the past activity of Sgr~A$^*$.
\end{abstract}


\section{Introduction}
\label{Intro}

Situated at a distance of $\sim$~8~kpc, the central region of our Galaxy hosts Sgr~A$^*$, the closest-to-Earth supermassive, 
4.3$\times$10$^6$~M$_\odot$, black hole \citep{Ghez2008,Gillessen2009}. Around the potential well of Sgr~A$^*$ is a 
concentration of active star formation sites and gigantic molecular clouds \citep{Yusef2008,Yusef2009} that makes the 
Galactic Centre (GC) of the Milky Way an excellent site for testing theories against observations in the context of an 
accreting black hole. It appears, however, that the present accretion rate of Sgr~A$^*$ is extremely small, about 
10$^{-8}$ M$_{\odot}$.y$^{-1}$ near the event horizon \citep{Baganoff2003}, so the X-ray luminosity is only about
$L_X~\sim$~2~$\times$~10$^{33}$~ergs.s$^{-1}$ \citep{Baganoff2001,Quataert2002}. 

Such quiescence is in disagreement with past X-ray observations of the Eastern massive molecular cloud Sgr~B2 that revealed 
a very steep spectrum with a strong emission line at 6.4~keV related to iron fluorescence, suggesting that part of the diffuse 
emission of Sgr~B2 is due to reprocessing \citep{Sunyaev1993,Koyama1996}. The lack of nearby X-ray sources, bright enough 
to account for the spectrum of Sgr~B2, lead \citet{Sunyaev1993} to classify Sgr~B2 as a reflection nebula, shining the 
reprocessed emission from a past Sgr~A$^*$ outburst ($L_X >$~10$^{39}$~erg.s$^{-1}$).
From geometrical considerations, \citet{Sunyaev1998} and \citet{Murakami2000} postulated that Sgr~A$^*$ underwent a 
flaring period that illuminated the reflection nebula, a hypothesis consolidated by the discovery of a similar behavior in 
the Western Sgr~C complex \citep{Murakami2001}. From a long-term time variability study, \citet{Inui2009} inferred that 
the outburst happened $\sim$~300 years ago, a conclusion shared by \citet{Ponti2010} who found that Sgr~A$^*$ was active 
$\sim$~400 years ago and again about 100 years ago. The estimated duration of the flare depends on the spatial location of 
the reflector which, unfortunately, cannot be properly constrained using X-ray spectroscopy or timing analyzes.

To overcome this difficulty, \citet{Churazov2002} proposed X-ray polarization as a mechanism to: 1) probe the flaring theory
as the resulting, reprocessed X-ray should be polarized and 2) estimate the three-dimensional location of the clouds.
Using a single cloud of radius 10~pc, with Thomson optical depth of 0.5 and filled with neutral, solar abundance matter, 
they show that the reflected radiation should be highly polarized ($>$~30 per cent) with a direction of polarization normal to
the scattering plane. Using a simple model, \citet{Churazov2002} proved that any detection of polarized X-ray emission 
from the reflection nebulae around Sgr~A$^*$ would bring constraints on the morphology and position of the scattering 
clouds, the duration of the flare and the location of the illuminating region. Yet, to be compared with future 
X-ray polarimetric observations, GC modeling needs further refinement. In particular, radiative coupling between 
the reflection nebulae and multiple scattering must be taken into account for time-delay, particularly in the case of 
short duration flares \citep{Sunyaev1998,Churazov2002}. Moreover, the dense and warm environment around the central few parsecs 
around Sgr~A$^*$ is likely to alter the polarization vector of radiation reaching the molecular clouds. The presence of the 
most massive component of the central molecular zone along the Galactic plane (sometimes referred as the ``disc population'', 
\citealt{Heiligman1987,Bally1988}), forming dust lanes along the observer's line-of-sight, is an additional component 
to the system that can alter polarization results by absorption and reprocessing. 
 
Intending to provide a more accurate estimation of the X-ray polarization signal that can emerge from multiple scattering, 
absorption and dilution from the source, we build a 2$^\circ~\times~$2$^\circ$ ($\sim$~288$~\times~$288~pc; at 8.5~kpc 
1$^{\prime\prime}$ $\approx$ 0.04~pc) model of the GC in Sect.~\ref{Model}. 
We analyze the resulting 8 -- 35~keV polarization map in Sect.~\ref{Results}, estimating both integrated and localized polarized 
X-ray emission. We extend the morphological parameterization in order to explore the influence of the location of the reprocessing 
components onto polarization in Sect.~\ref{NewParameters} and investigate the polarization detectability of Sgr~B2 and Sgr~C in 
Sect.~\ref{Detectability}. We discuss our work and draw conclusions in the last sections of this paper.

\section{Polarized emission from the Galactic Centre}
\label{Main}

The complex emitting and scattering environment of Sgr~A$^*$ remains enigmatic in terms of morphology. Constraints on the 
central hundred parsecs derived by spectroscopy and velocity measurement in the radio (e.g. \citealt{Tsuboi1999}), 
infrared (e.g. \citealt{Molinari2010}) and X-ray/soft-$\gamma$ domains (e.g. \citealt{Sunyaev1991}) helped to estimate 
the three-dimensional gas distribution. Using the geometrical constraints obtained so far, we now investigate the resulting 
polarization signatures from the GC using X-ray polarimetry. Knowing that past X-ray observations \citep{Koyama1986,Koyama1989,Sidoli1999}
have revealed the presence of a diffuse plasma emission toward the GC that could dilute the polarization signal below 7~keV 
\citep{Mewe1999,Liedahl1999}, we limit the investigation to the soft X-ray band. We center our analysis in the 8 -- 35~keV energy 
domain to: 1) avoid most of the dilution by the GC plasma emission, 2) avoid local polarization dilution by the unpolarized 6.4~keV 
iron emission line and 7.1~keV edge structure \citep{Murakami2000}, and 3) extend the 2 -- 8~keV simulations achieved by 
\citet{Churazov2002}.

\subsection{Modeling X-ray polarization from the inner 100~pc}
\label{Model}

\begin{figure}
   \centering
      \includegraphics[trim = 0mm 0mm 0mm 0mm, clip, width=8cm]{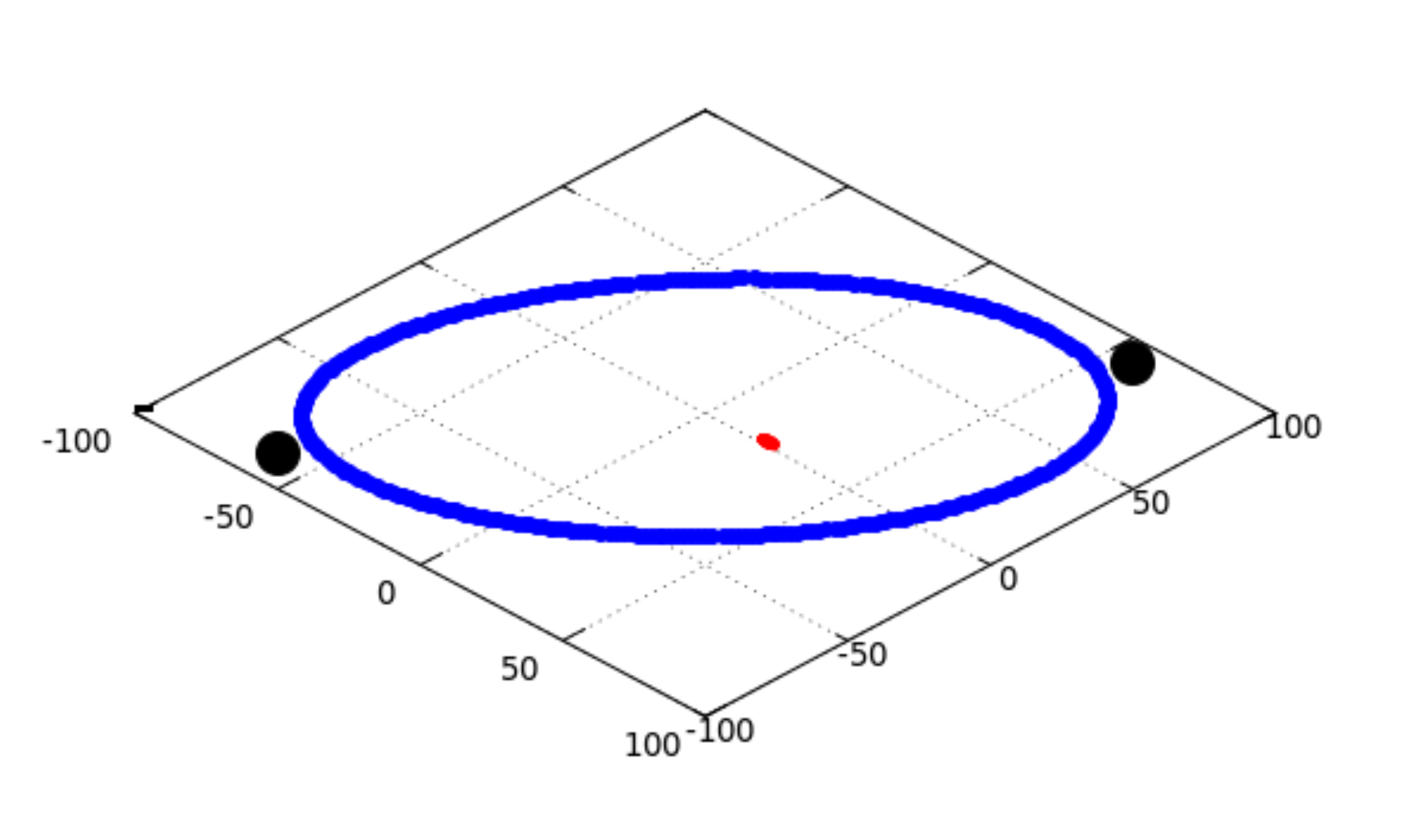}
      \includegraphics[trim = 0mm 0mm 0mm 0mm, clip, width=8cm]{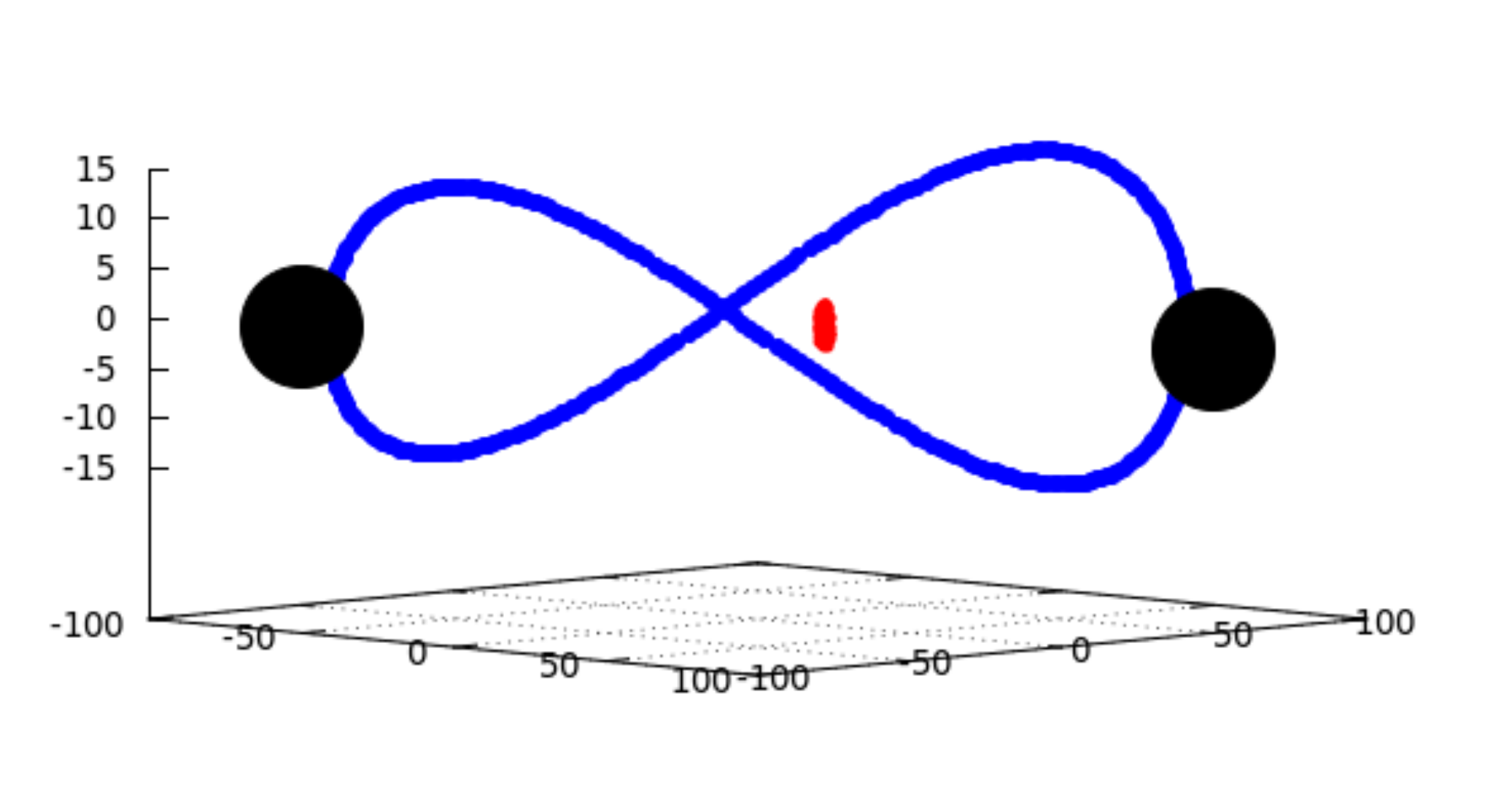}
      \caption{Sketch of the three-dimensional GC structure. The 100~pc elliptical twisted ring 
	       \citep{Molinari2011} is drawn in blue, the fragmented circumnuclear disc
	       surrounding Sgr~A$^*$ in red, and Sgr~B2 (left, Eastern cloud) and Sgr~C (right, Western \
	       cloud) in black. Axes are in parsec units and the sketch is roughly to scale. $Top$: 
	       pole-on view of the GC; $bottom$: edge-on (observed) view of the GC.}
     \label{Fig:Sketch}
\end{figure}

Polarization modeling was achieved using {\sc stokes} \citep{Goosmann2007,Marin2012a}, a Monte Carlo code that takes into 
account the polarized component of light. {\sc stokes} performs three-dimensional radiative transfer simulations within various 
geometries of reprocessing regions. Originally developed for optical/UV investigation, the code was extended to the X-ray range 
with the inclusion of inelastic scattering responsible for the Compton hump above 10 -- 20~keV, absorption, photo-ionisation and
recombination effects. Detailed information about {\sc stokes} calculation of X-ray polarization are given in \citet{Goosmann2010}.

In the context of the GC, unpolarized input photons are generated from a point-like, isotropic source that emits 8 -- 35~keV photons 
according to a power-law spectral energy distribution $F_{\rm *}~\propto~\nu^{-\alpha}$. The spectral index $\alpha$ is fixed to unity 
in order to match the photon index of Sgr~A$^*$ flaring events ($\sim$~2, \citealt{Porquet2003,Baganoff2003,Belanger2005,Neilsen2013,Barriere2014}) 
that can be representative of the past activity of the supermassive black hole. Finally, the source is displaced by $\sim$~22~pc in projection 
from the center of the model toward the Western galactic longitude to be consistent with the shifted gas distribution within the central 
molecular zone (CMZ, \citealt{Molinari2011}). 

In our model (a sketch is presented in Fig.~\ref{Fig:Sketch}.) the inner 5~pc around Sgr~A$^*$ is surrounded by a clumpy distribution of 
100 molecular gas spheres filled with cosmic abundance matter (hydrogen, helium, carbon, nitrogen, oxygen, neon, magnesium, silicate, sulfur 
and iron). The gas clouds, representative of the circumnuclear disc (CND), are arranged in a configuration similar to a fragmented disc tilted by 
20$^\circ$ with respect to the Galactic plane \citep{Ponti2013,Liu2014}. The clumps have an hydrogen column density $>~$10$^{25}$~atom.cm$^{-2}$ 
and an orbital velocity of 100~km.s$^{-1}$ \citep{Christopher2005}; the gas temperature is fixed to 200~K \citep{Mills2013}. 

At larger distances from Sgr~A$^*$ and the CND, we implement a cold, dusty structure as proposed by \citet{Molinari2011}. In their 
picture, a continuous chain of irregular clumps forms the CMZ in the shape of an elliptical twisted ring of molecular material. The 
$\infty$-shaped ring is parameterized according to \citet{Molinari2011}\footnote{The original formula by \citet{Molinari2011} contains a 
typo. One must read $z = z_0 \sin\nu_{\rm orb}(\theta_t - \theta_z)$ instead of $z = z_0 \sin\nu_{\rm orb}(\theta_p - \theta_z)$, with the 
definitions of $z, z_0, \nu_{\rm orb}, \theta_t, \theta_p$ and $\theta_z$ to be found in Sect.~3.1 of their paper.}. The resulting CMZ is 
composed of 300 spherical clouds with hydrogen column density 10$^{24}$~atom.cm$^{-2}$ \citep{Ponti2013}, gas temperature 80~K \citep{Morris1983} 
and orbital velocity 80~km.s$^{-1}$ \citep{Molinari2011,Ponti2013}. On the plane-of-the-sky, the $\infty$-shaped ring extends up to $\sim$~100~pc 
in longitude and $\sim$~25~pc in latitude. 

We include into our model the two bright, complex reflection nebulae Sgr~B2 and Sgr~C. Located in the Eastern CMZ, Sgr~B2 is the largest 
and heaviest molecular cloud within the inner 100 parsecs around Sgr~A$^*$ \citep{Murakami2001} and thus a prime target for scattering 
induced polarization. Sgr~C is a complex structure composed of, at least, three separate components distant from each other,
and holds the largest star forming region in the Western CMZ \citep{Kendrew2013}. Sgr~B2 and Sgr~C are located at the two extrema of the 
twisted elliptical ring and, according to the dependence on the cosine square of Thomson scattering, are likely to be the best targets for 
future X-ray polarimetric measurement. Sgr~B2 is defined as a spherical cloud with an overall radius 7~pc, hydrogen column density 
8$\times$10$^{23}$~atom.cm$^{-2}$ and orbital velocity 60~km.s$^{-1}$ \citep{Ponti2010}. Sgr~C has an overall radius of 5~pc \citep{Ryu2013}, a hydrogen 
column density equals to 8$\times$10$^{22}$~atom.cm$^{-2}$ \citep{Kendrew2013} and we fix its orbital velocity to 60~km.s$^{-1}$ (similar to Sgr~B2). 
Both clouds are filled with neutral, cosmic abundance matter and are situated at $\sim$~100~pc from the center of the model. However, 
from very-long-baseline interferometry (VLBI) and very long baseline array (VLBA) trigonometric parallax arguments, \citet{Reid2009} 
situate Sgr~B2 in the front of the Galactic plane. To investigate the importance of the position of the scattering nebulae, we thus 
shift Sgr~B2 by $\sim$~10~pc toward the observer while Sgr~C remains closer to the Western extrema. 

Polarization being sensitive to any departure from symmetry, the combination of the structure proposed by \citet{Molinari2011}, the off-center 
irradiating source, and the spatial shift of Sgr~B2 (and to a lesser extent the one of Sgr~C) are expected to enhance the production of 
spatially localized polarization.

\subsection{Polarization results}
\label{Results}

\begin{figure}
   \centering
      \includegraphics[trim = 5mm 5mm 0mm 10mm, clip, width=8cm]{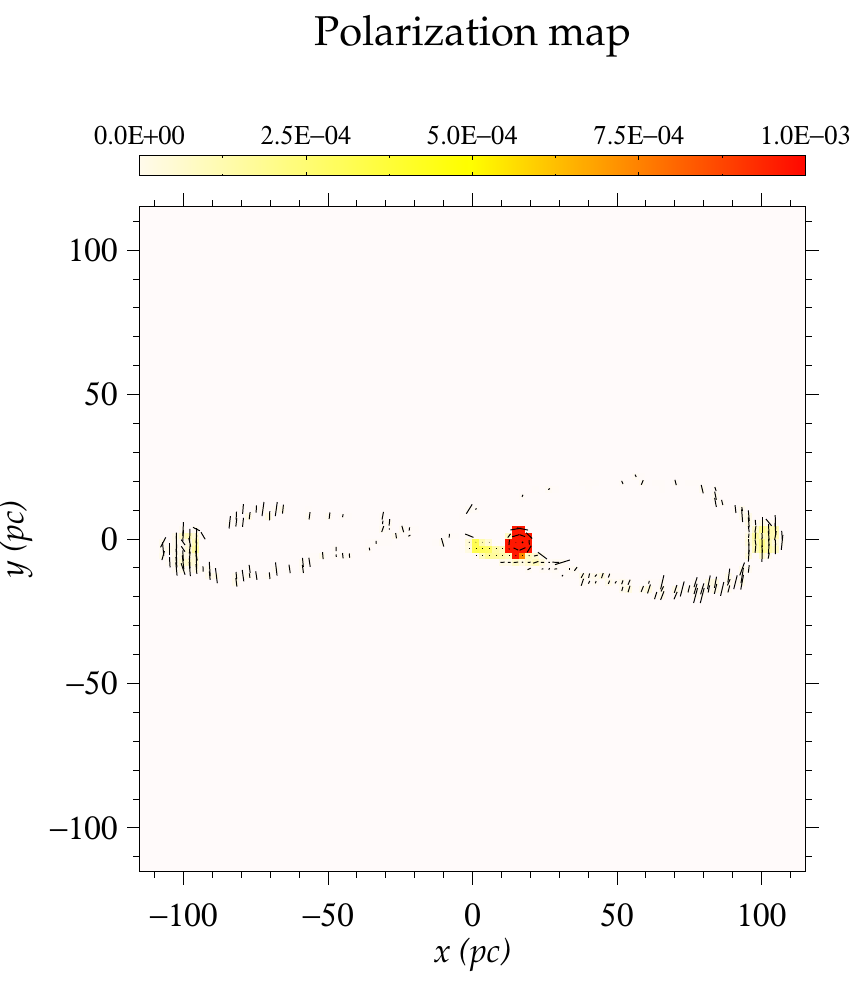}
      \caption{Integrated 8 -- 35~keV model image of the polarized flux, $PF/F_{\rm *}$,
	for the 2$^\circ~\times~$2$^\circ$ region around the GC. $PF/F_{\rm *}$ is 
	color-coded, with the color scale shown on top of the image (in arbitrary units).
	$P$ and $\psi$ are represented by black bars drawn in the center 
	of each spatial bin. A vertical bar indicates a polarization angle of 
	$\psi$~=~90$^\circ$ and a horizontal bar stands for an angle
	of $\psi$~=~0$^\circ$. The length of the bar is proportional to 
	$P$.}
     \label{Fig:Map}
\end{figure}

We sampled a total of 10$^9$ photons and present in Fig.~\ref{Fig:Map} an integrated, 8 -- 35~keV, polarization map of the GC. The spatial 
resolution is set to 100$~\times~$100~bins for the longitudinal and latitudinal axes, so that the photon flux is divided into 10$^4$ pixels. 
Each of these pixels is labeled by the position offset in parsecs and stores the four Stokes parameters of the 8 -- 35~keV photons. The resulting 
polarized flux ($PF/F_{\rm *}$, i.e. intensity times polarization degree) is normalized to the central flux $F_{\rm *}$ that is emitted into the same 
viewing direction, and is color-coded. The polarization degree $P$, ranging from 0 per cent (unpolarized) to 100 per cent (fully polarized), and the polarization 
position angle $\psi$ are represented by black bars drawn in the center of each spatial bin. A vertical bar indicates a polarization position angle 
$\psi$~=~90$^\circ$ (perpendicular to the projected vertical axis of the model) and a horizontal bar stands for an angle $\psi$~=~0$^\circ$ 
(parallel to the projected vertical axis). The length of the bar is proportional to $P$.

~\

\begin{table}
  \centering
  {
   \footnotesize
   \begin{tabular}{|c|c|c|}
   \hline
      {\bf Region}		& {\bf $P$ [\%]}	& {\bf $\psi$ [$^\circ$]}\\
   \hline
      GC (integrated)   	& 0.9			& -22.8\\
      CND    			& 1.0 	         	& -20.5\\
      Sgr~B2    		& 66.5            	& 88.1\\      
      Sgr~C    			& 47.8           	& 89.5\\
   \hline
   \end{tabular}
  }
  \caption{Estimated 8 -- 35~keV polarization degree $P$ and polarization angle $\psi$ of the 
	   2$^\circ~\times~$2$^\circ$ GC and individual reprocessing regions. Polarization angles 
	   are defined with respect to the projected vertical axis of the system.}
  \label{Pol_res}
\end{table}

The polarized flux, $PF/F_{\rm *}$, traces the overall shape of the GC, emphasizing the $\infty$-shaped CMZ. $PF/F_{\rm *}$ 
reaches a maximum at the location of the CND surrounding Sgr~A$^*$, where the combination of the flux from the emitting source 
and the polarization of reprocessed photons by neutral material increases the polarized flux. However, the local polarization degree 
is rather low, $\sim$~1.0 per cent, when integrated over the CND (see Tab.~\ref{Pol_res}). This is due to the disc-like geometry of 
the scattering material that is seen nearly pole-on by the observer. Thomson scattering-induced polarization being small 
for forward and backward scattering, the morphology of the CND thus favors low $P$. Increasing the filling factor of the 
CND or replacing it by a uniform disc has very little impact on $P$. The polarization position angle associated with the CND 
is not null ($\psi$ = -20.5$^\circ$) as the system is inclined by 20$^\circ$ with respect to the Galactic plane \citep{Ponti2013}.

Sgr~B2 and Sgr~C show the secondary, brightest polarized flux knots of the map. Polarization by electron scattering being the 
most efficient for orthogonal scattering angles, we find $P$ = 66.5 per cent for Sgr~B2 and $P$ = 47.8 per cent for Sgr~C. The difference 
in $P$ is due to the spatial location of the cloud with respect to Sgr~A$^*$ and the resulting asymmetry produced by this setup. 
Both the clouds present a polarization position angle roughly perpendicular to the vertical axis of the model, as predicted by 
\citet{Churazov2002}. By measuring the angle between the emitting source, the cloud and the observer, X-ray polarimetry could
properly reconstruct the position of the primary emitting source.

The CMZ polarization traces the shape of the structure, with partial disappearance of $P$ due to low polarization degrees
induced by forward and backward scattering, and dilution from unpolarized radiation from the source. $P$ varies for each 
CMZ cloud and reaches a maximum for the Eastern and Western apoapse sections of the elliptical twisted ring, where the CMZ
mixes with the giant molecular clouds. It is most likely that a large fraction of the $\infty$-shaped ring will be 
diluted by background, unpolarized emission from both plasma emission and Sgr~A$^*$, and thus be undetectable.

Finally, when integrating the whole 2$^\circ~\times~$2$^\circ$ GC polarized emission, the model produces a net polarization
degree of 0.9 per cent associated with a polarization position angle of -22.8$^\circ$. The combined emission from Sgr~A$^*$ and the CND 
thus dominates the whole polarization picture.

\subsection{Impact of the location of components}
\label{NewParameters}

Despite constraints brought by multi-wavelength observations \citep[e.g.][]{Eckart2008,Law2008}, the GC remains puzzling in terms 
of gas morphology, composition and location. The three-dimensional parameterization modeled in Sect.~\ref{Results} mainly relies
on projected radial distances with respect to the Galactic plane. As noted by \citet{Reid2009}, the actual location of the reflecting
nebulae can be shifted from this plane but still conserve the same radial projections. Polarimetry being sensitive to the 
geometry of the scattering system, we now depart from the assumed GC parameters listed in Sect~\ref{Model} and test different 
locations of Sgr~B2, Sgr~C and Sgr~A$^*$ (coupled to the CND).

~\

\begin{figure}
   \centering
      \includegraphics[width=8cm]{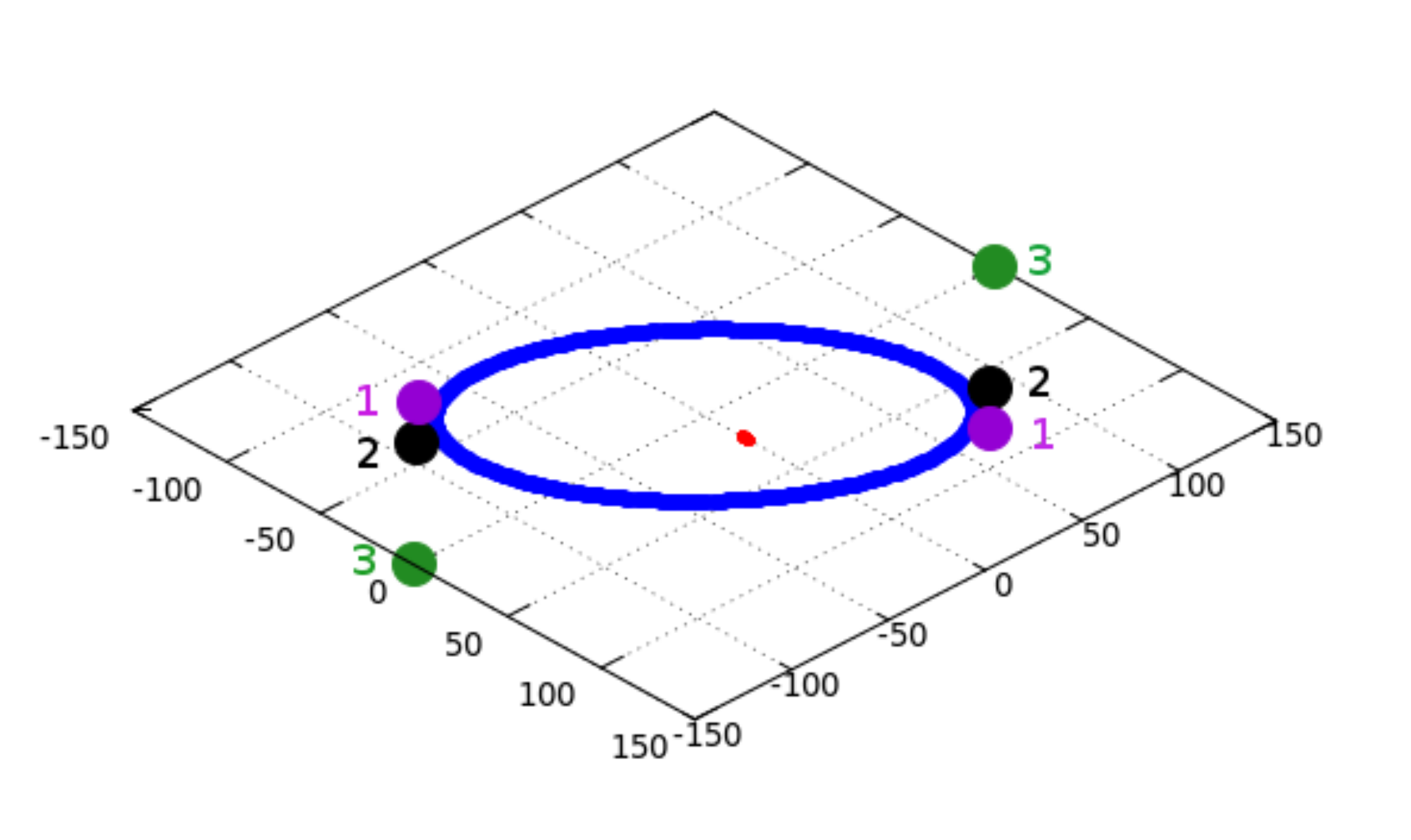}
      \caption{Investigating different locations of Sgr~B2 and Sgr~C
	        that produce the same projected distances on the Galactic plane.	
	        The sketch is roughly to scale. $Legend$: clouds situated on the 
		Galactic plane are shown in violet, clouds at the distances
		fixed in Sect.~\ref{Model} in black and clouds situated at 
		$\sim$~100~pc from the Galactic plane in green. The same colors 
		are used in Tab.~\ref{Tab:DiffClouds} to identify 
		$P$ and $\psi$ for each case.}
     \label{Fig:DiffClouds}
\end{figure}

We first explore the impact of different locations of Sgr~B2 and Sgr~C with respect to the Galactic plane, conserving 
a projected distance similar to the one derived by \citet{Molinari2011}. For an observer situated on Earth, the three 
models summarized in Fig.~\ref{Fig:DiffClouds} (seen from the pole) produce the same edge-on GC image. Similarly to our 
previous investigation, we spatially integrate the 8 -- 35~keV net polarization of the three main scattering regions
(CND, Sgr~B2 and Sgr~C) and we also evaluate the resulting polarization from the whole 2$^\circ~\times~$2$^\circ$ GC
in Tab.~\ref{Tab:DiffClouds}. The first model, with violet clouds, reproduces the extreme case where the two reflection 
nebulae lie at the apoapsis of the elliptical twisted ring of gas. As the scattering angle between the source, the 
reprocessing material and the observer is nearly orthogonal, $P$ is maximum: 71.0 per cent for Sgr~B2 and 64.0 per cent for Sgr~C, 
both associated with a polarization position angle of $\sim$~90$^\circ$. When the reflection nebulae depart from the Galactic 
plane, $P$ decreases as the scattering angle becomes narrower. For distant, $\sim$~100~pc, scattering clouds (represented in 
green), $P$ equals to 38.6 per cent for Sgr~B2 and 16.0 per cent for Sgr~C, with $\psi$ remaining constant. The spatial location of Sgr~B2 
and Sgr~C can thus be estimated from a measurement of the net polarization degree, as $P$ decreases with the distance of the 
clouds from the Galactic plane. Independently of the position of the reflecting nebulae, $P_{\rm CND}$ and $\psi_{\rm CND}$ 
remain constant for the three models, still dominating the integrated GC picture despite the large amount of polarization 
produced by the scattering nebulae when the clouds are on the Galactic plane.

\begin{table}
  \centering
  {
   \begin{tabular}{c|c|c|c|c|c|c|}
      \hline
      {\bf Region} 	& \multicolumn{3}{c|}{{\bf $P$ [\%]}} & \multicolumn{3}{c|}{{\bf $\psi$ [$^\circ$]}}\\
      ~& \textcolor{BlueViolet}{M$_1$} & \textcolor{black}{M$_2$} & \textcolor{ForestGreen}{M$_3$} & \textcolor{BlueViolet}{M$_1$} & \textcolor{black}{M$_2$} & \textcolor{ForestGreen}{M$_3$}\\ 
      \hline
      GC 		& \textcolor{BlueViolet}{0.9}	& \textcolor{black}{0.9}	& \textcolor{ForestGreen}{1.0}		& \textcolor{BlueViolet}{-22.0}& \textcolor{black}{-22.8}	& \textcolor{ForestGreen}{-21.4}\\
      CND    		& \textcolor{BlueViolet}{1.0}	& \textcolor{black}{1.0}	& \textcolor{ForestGreen}{1.0}         & \textcolor{BlueViolet}{-18.7}& \textcolor{black}{-20.5}	& \textcolor{ForestGreen}{-19.9}\\
      Sgr~B2    	& \textcolor{BlueViolet}{71.0}	& \textcolor{black}{66.5}	& \textcolor{ForestGreen}{38.6}        & \textcolor{BlueViolet}{91.5}	& \textcolor{black}{89.1}	& \textcolor{ForestGreen}{91.3}\\      
      Sgr~C    		& \textcolor{BlueViolet}{64.0}	& \textcolor{black}{47.8}	& \textcolor{ForestGreen}{16.0}        & \textcolor{BlueViolet}{90.4}	& \textcolor{black}{89.5}	& \textcolor{ForestGreen}{89.6}\\
      \hline
  \end{tabular}
  }  
  \caption{Estimated 8 -- 35~keV polarization degree $P$ and polarization angle $\psi$ of the integrated 
	   2$^\circ~\times~$2$^\circ$ and individual reprocessing regions. M$_{1,2,3}$ and the colors 
	   refer to the three different radial positions of the colouds presented in Fig~\ref{Fig:DiffClouds}.}
  \label{Tab:DiffClouds}
\end{table}

~\

\begin{figure}
   \centering
      \includegraphics[width=8cm]{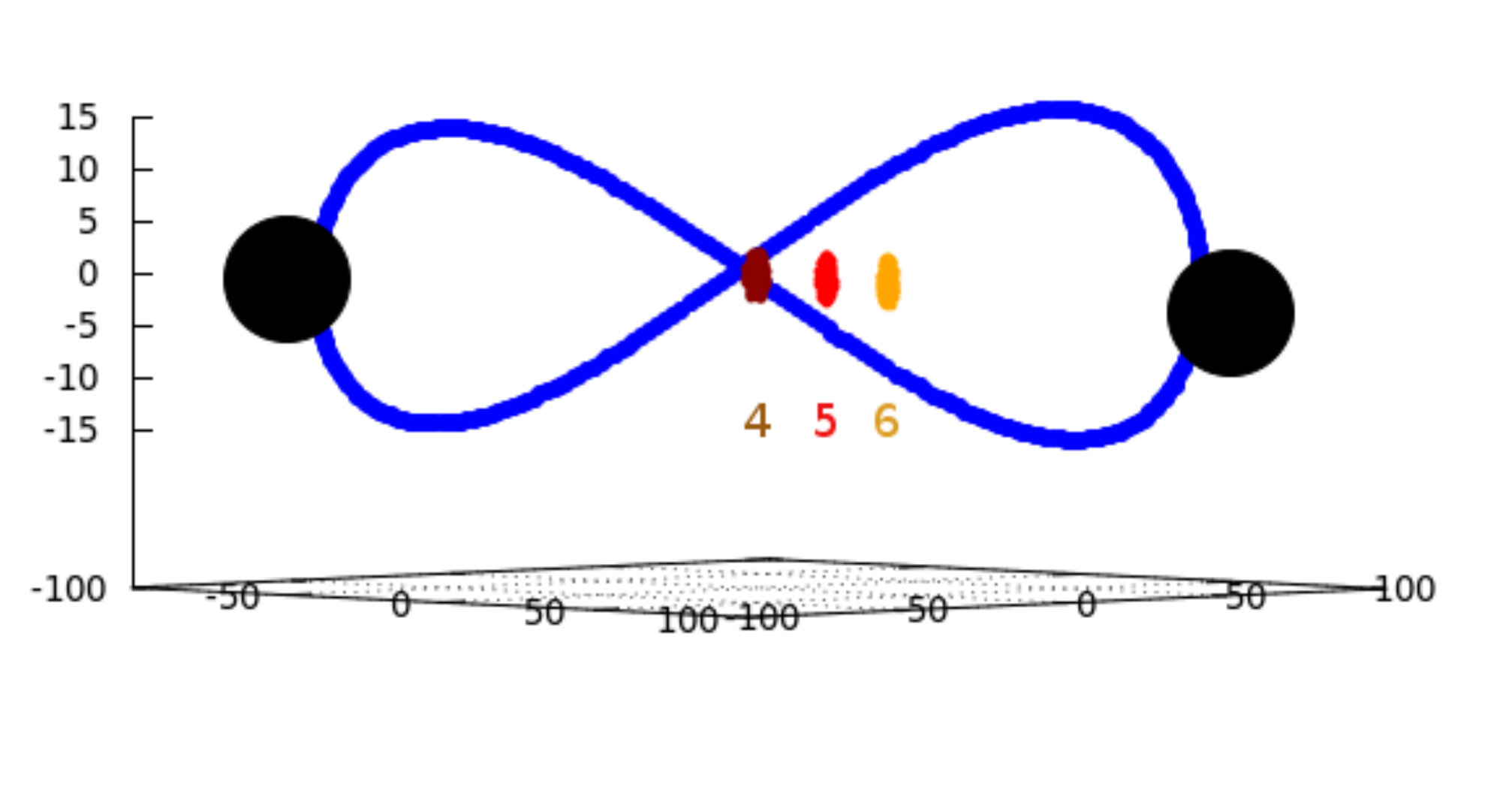}
      \caption{Investigating different projected distances for Sgr~A$^*$ / CND
		on the Galactic plane. The sketch is roughly to scale. 
		$Legend$: $\sim$~0~pc projected distance from the center of the 
		$\infty$-shaped ring in brown, distance fixed in Sect.~\ref{Model} 
		($\sim$~22~pc) in red and $\sim$~42~pc projected distance in orange. 
		The same colors are used in Tab.~\ref{Tab:DiffSgrAstar} to 
		identify $P$ and $\psi$ for each case.}
     \label{Fig:DiffSgrAstar}
\end{figure}

We then investigate the impact of the location of the central supermassive black hole and its surrounding CND on the 
total GC polarization. In Fig.~\ref{Fig:DiffSgrAstar}, we show three possible locations of Sgr~A$^*$, paired with the
CND, which give a projected distance of $\sim$~0~pc from the center of the model (in brown), $\sim$~22~pc (in red, such as 
in Sect.~\ref{Model}) and $\sim$~42~pc. Results for the 8 -- 35~keV integrated polarization are given in Tab.~\ref{Tab:DiffSgrAstar}.
Independent of the position of Sgr~A$^*$, the surrounding cloudlet distribution that forms the CND produces a constant
$\sim$~1.0 per cent polarization degree since the composition and morphology of the CND enclosing Sgr~A$^*$ did not vary. However, 
the polarization position angle differs between the model where the emitting source is situated at the center of the 
structure and models with shifted locations of Sgr~A$^*$. The difference is due to the $\infty$-shaped CMZ dust lane
at the forefront of model, partially covering the central supermassive black hole in the first scenario. To reach the observer,
radiation has to pass through this molecular strip, undergoing scattering events that alter its polarization position
angle and marginally diminish the net polarization degree due to extra absorption. Similarly to Fig.~\ref{Fig:Map}
and Fig.~\ref{Fig:DiffClouds}, the integrated GC polarization is dominated by the reprocessed emission from the Sgr~A$^*$ 
/ CND duo. $P_{\rm Sgr~B2}$ and $P_{\rm Sgr~C}$ also show variations related to the location of the emitting source. 
While the polarization degree of $P_{\rm Sgr~B2}$ increases with the (projected) departure of the emitting source from the 
center of the model, $P_{\rm Sgr~C}$ diminishes. The source of unpolarized photons getting closer to $P_{\rm Sgr~C}$, it 
enhances the local dilution of polarization and thus reduces the net Western $P$. The opposite effect occurs on the 
Eastern side of the model. The resulting polarization is still high ($>$~35 per cent) and both the reflection nebulae produce 
a perpendicular polarization position angle, since the Sgr~A$^*$ / CND duo moved along the Galactic longitude only.

\begin{table}
  \centering
  {
   \begin{tabular}{c|c|c|c|c|c|c|}
      \hline
      {\bf Region} 	& \multicolumn{3}{c|}{{\bf $P$ [\%]}} & \multicolumn{3}{c|}{{\bf $\psi$ [$^\circ$]}}\\
      ~& \textcolor{Brown}{M$_4$} & \textcolor{red}{M$_5$} & \textcolor{Orange}{M$_6$} & \textcolor{Brown}{M$_4$} & \textcolor{red}{M$_5$} & \textcolor{Orange}{M$_6$}\\ 
      \hline
      GC 		& \textcolor{Brown}{0.7}	& \textcolor{red}{0.9}	& \textcolor{Orange}{0.9}	   & \textcolor{Brown}{-39.8}  & \textcolor{red}{-23.4}	& \textcolor{Orange}{-21.8}\\
      CND    		& \textcolor{Brown}{0.7}	& \textcolor{red}{1.0}	& \textcolor{Orange}{0.9}         & \textcolor{Brown}{-36.0}  & \textcolor{red}{-21.3}	& \textcolor{Orange}{-20.3}\\
      Sgr~B2    	& \textcolor{Brown}{58.6}	& \textcolor{red}{67.0}	& \textcolor{Orange}{73.8}        & \textcolor{Brown}{90.0}	& \textcolor{red}{89.0}		& \textcolor{Orange}{89.0}\\      
      Sgr~C    		& \textcolor{Brown}{53.0}	& \textcolor{red}{44.5}	& \textcolor{Orange}{37.3}        & \textcolor{Brown}{93.0}	& \textcolor{red}{88.0}		& \textcolor{Orange}{91.7}\\
      \hline
  \end{tabular}
  }  
  \caption{Estimated 8 -- 35~keV polarization degree $P$ and polarization angle $\psi$ of the integrated 
	   2$^\circ~\times~$2$^\circ$ and individual reprocessing regions. M$_{4,5,6}$ and the colors 
	   refer to the three different projected distances of Sgr~A$^*$ / CND presented in 
	   Fig~\ref{Fig:DiffSgrAstar}.}
  \label{Tab:DiffSgrAstar}
\end{table}

\subsection{Detectability with available technology}
\label{Detectability}

The technology enabling the measurement of celestial X-ray polarization has greatly evolved since the 1970s, when the {\it 8th 
Orbiting Solar Observatory} ({\it OSO-8}) carried the last flying non-solar X-ray polarimeter \citep{Weisskopf1978}. The new generation 
of Gas Pixel Detector (GPD), based on the photoelectric effect \citep{Costa2001}, overcomes the sensitivity limits of past X-ray 
polarimeters based on graphite-crystal Bragg diffraction at 45$^\circ$ or Thomson/Compton scattering, and is thus a new and 
technologically-ready instrument to probe the GC \citep{Tagliaferri2012a,Soffitta2013}.

Results from Sect.~\ref{Results} and \ref{NewParameters} showed that Sgr~B2 and Sgr~C are prime candidates for polarization observations 
as they are expected to produce a significant amount of polarization ($\gg$~10 per cent). Hence, to strengthen the case of X-ray polarimetry
for the future of X-ray astronomy, we now investigate their detectability by exploring the minimum detectable polarization (MDP) that a 
past X-ray polarimetric project could have reached if it had been selected for launch. We focus on {\it NHXM}, the {\it New Hard X-ray Mission} 
\citep{Tagliaferri2012a,Tagliaferri2012b}, an M-class satellite designed to carry a 2 -- 10~keV low-energy polarimeter (LEP) and a 6 -- 35~keV 
medium-energy polarimeter (MEP). To avoid dilution by the GC plasma emission, the MEP is then the most suited instrument to measure 
polarization from Sgr~B2 and Sgr~C. The angular resolution (about 20 arcsec) would have been able to spatially resolve the two 
reflection nebulae, discriminating them from neighborhood sources and potentially enabling the investigation of stratified light echoes from 
the past activity of Sgr~A$^*$.

To calculate the 99 per cent confidence level MDP of the medium/hard X-ray instrument (see \citealt{Marin2012b,Strohmayer2013}), we derived the 
flux of Sgr~B2 in the MEP band (excluding the iron lines energies where the signal is diluted by unpolarized fluorescence emission) from the 
20 -- 60~keV energy band \citep{Terrier2010}. The power-law index \citep{Murakami2000,Sidoli2001,Terrier2010} being close to the one of the Crab 
($\Gamma~\sim$~2), in first approximation we can assume similar fluxes (in mCrab) in the MEP energy band if we neglect absorption below 8~keV. 
No observation of Sgr~C has been made above 10~keV yet, but its soft X-ray continuum has been fitted with a $\Gamma~\sim$~2 power law 
\citep{Murakami2001,Takagi2002}, similar to Sgr~B2. Therefore, we assume for Sgr~C a flux in the MEP energy band which is just scaled with 
respect to that of Sgr~B2 by the same factor which is observed in the soft X-rays, that is, a factor 3 fainter \citep{Murakami2000}.
As the count rates from the molecular clouds are faint, we integrate the flux in the largest possible energy range to provide a representative 
value of the 8 -- 35~keV MDP for a 500~ks observation (further details are provided in \citealt{Marin2013}). Estimations on the detected polarization 
P$_{\rm detect.}$ and the error on $\psi$ are calculated from Monte Carlo simulations, such as presented in \citet{Dovciak2011}, and are derived 
from the generation of test modulation curves. The expected initial polarization P$_{\rm source}$ is taken from our modeling of the GC examined in 
Sect.~\ref{Results}.

Results are presented in Tab.~\ref{MDP_est}. We find that both reflecting nebulae are detectable using a 500~ks observation with {\it NHXM} 
(MDP~$<$~P$_{\rm source}$). Longer exposure time would help to reach lower MDP values and increase the statistic that becomes relevant to measure 
the polarization angle, but Tab.~\ref{MDP_est} shows that a medium-sized polarimetric mission would already be able to measure the polarization 
emerging from Sgr~B2 and Sgr~C with great precision. Errors on $\psi$ being marginal, the detection of polarization position angle normal to
the scattering plane would be unambiguous. Hence, since unpolarized emission from Galactic plasma emission in the soft X-ray band should weaken 
polarization detection, a mission equipped with a broadband polarimeter is the most suited to provide the cleaner measurement.

\begin{table}
  \centering
  {
   \footnotesize
   \begin{tabular}{|c|c|c|c|c}
   \hline
      {\bf Region}	& {\bf P$_{\rm source}$}	& {\bf MDP} 	& {\bf P$_{\rm detect.}$} 	& {\bf Error on $\psi$}\\
   \hline
      Sgr~B2    	& 66.5~\%			& 7.7~\%	& 66.7$~\pm~$1.8~\%		& 0.63$^\circ$\\      
      Sgr~C    		& 47.8~\%        		& 4.5~\%	& 48.7$~\pm~$4.6~\%		& 1.52$^\circ$\\
   \hline
   \end{tabular}
  }
  \caption{Minimum detectable polarization (MDP) of the reflection nebulae with a 500~ks observation with 
	   the MEP on board of {\it MHXM}. Predictions on the detected polarization P$_{\rm detect.}$
	   and the error on $\psi$ are calculated from Monte Carlo simulations.}
  \label{MDP_est}
\end{table}

\section{Discussion}
\label{Discussion}

\subsection{Additional targets}
\label{Degeneracies}

Our simulations showed that the three-dimensional position of the reprocessing Sgr~B2 and Sgr~C clouds can be 
estimated using X-ray polarization measurement above 7~keV. Clouds situated at large distances from the Galactic plane will 
produce lower polarization degrees as the angle between the source, the scattering medium and the observer will depart 
from orthogonality. $P$ is thus a function of the position of the emitting source. Targeting a unique reflection nebula will 
indeed prove or reject the outburst scenario by measuring $\psi$ \citep{Churazov2002}, but it will be necessary to observe at 
least two nebulae to constrain the three-dimensional location of Sgr~A$^*$. From them it will become easier to reconstruct the
layout of circumnuclear clouds with respect to the supermassive black hole in the GC. However, Tab.~\ref{Tab:DiffClouds} 
showed that the polarization degree of clouds either at the rear or in front of the Galactic plane decreases as they 
depart from the plane-of-the-sky. From only two measurements it might be difficult to decide if the clouds are behind or 
in front of this plane. To break potential degeneracies it will be necessary to measure as many reflection nebulae as possible 
in order to numerically reconstruct a three-dimensional structure of the GC that coherently reproduces each polarization 
percentage detected.

The dense environment of the GC provides several other potential targets in which hard X-ray continuum and Fe~K$\alpha$ emission 
have been found, spatially coincident with GC molecular clouds: Sgr~B1 \citep{Koyama2007}, M0.74-0.09 \citep{Koyama2007}, 
G0.11-0.11 \citep{Ponti2010}, M0.74-0.09 \citep{Nobukawa2011}, M1 \citep{Ponti2010}, M2 \citep{Ponti2010} or the molecular 
complex called the Bridge \citep{Gusten1980,Bamba2002}. Measuring $P$ and $\psi$ from such a collection of reflection nebulae 
will help us to determine their location with respect to Sgr~A$^*$ and test if the central supermassive black hole was 
the unique, faded illuminating source. One must be cautious though, as some of these scattering clouds present the same 
characteristic as Sgr~B2: a decay of their Fe~K$\alpha$ emission \citep{Koyama2008,Inui2009,Terrier2010}. Flux diminution 
is consistent with the reflection mechanism as light needs time to reach the core of the molecular cloud, facing an 
increasing molecular density and thus higher absorption probability. Sgr~B2 and Sgr~C being the brightest sources
among the reflection nebulae, they remain the best targets for future X-ray polarization measurement, despite 
the fact that the duty cycle of the GC accretion activity and the flux of molecular clouds in the next decade (the time frame 
of a possible mission able to measure the polarization) are an open question. Since the efficiency of a detection relies on the 
luminosity of the source, the MDP can become too large for several objects, such as Sgr~B1 and M0.74-0.09, both fainter 
than Sgr~B2 and also exhibiting temporal decreasing intensity \citep{Koyama2007}.

\subsection{Alternative scenario for the origin of X-ray emission from Sgr~B2 and Sgr~C}
\label{LECRE}

X-ray polarimetry can be a powerful tool to examine other scenarios in which the power-law continuum and the 6.4~keV iron 
feature detected in Sgr~B2, Sgr~C and other sources, such as the Galactic ridge \citep{Valinia2000} and G0.13--0.13 \citep{Yusef2002}, 
are produced by low-energy cosmic-ray electrons rather than by Thomson/Compton scattering \citep{Valinia2000,Yusef2002,Dogiel2009,Yusef2013}. 
In this model \citep{Valinia2000,Yusef2002}, nonthermal X-ray emission arises from the interaction of fast electrons ($E~<~$1~MeV)
with the GC molecular clouds. Emission lines result from the filling of inner-shell vacancies produced by the low-energy cosmic-ray 
electrons traveling in the atomic mixture while the power-law continuum results from bremsstrahlung processes. Such a mechanism 
provides fits of similar quality when compared to neutral reflection from a flaring source \citep{Yusef2002} but polarization can 
discriminate between the two interpretations. 

Bremsstrahlung radiation is generally polarized with its electric vector perpendicular to the plane of interaction. In the case of 
thermal bremsstrahlung, the planes of interaction are randomly distributed, resulting in null net polarization. Bremsstrahlung emission 
can be polarized only for an anisotropic distribution of electrons. However, very little is known about the atomic and molecular 
distribution in the reflection nebulae. Looking at the central 12~pc of the Sgr~B2 molecular cloud in the 3~mm band, \citet{Jones2008} 
identified at least seven distinct molecular features concentrated at various positions in the cloud complex. The structure of Sgr~B2 
was previously estimated with $J~=~$1$~\rightarrow~$0 transition of $^{13}$CO and C$^{18}$O by \citet{Lis1989}, who also found a 
stratification of the cloud, with an increase of $^{13}$CO optical depth at the center of the cloud. Hints tend to point toward an 
asymmetric distribution of matter within the molecular clouds and thus should lead to polarized bremsstrahlung emission, which 
remains to be estimated.

\section{Conclusions}
\label{Conclusion}

The simulations undertaken in this paper consolidate the results from the toy-model investigated by \citet{Churazov2002}
in the context of a future X-ray polarimetric observations. The 2$^\circ~\times~$2$^\circ$ Galactic inner region is expected 
to be dominated by large polarized fluxes originating from the CND surrounding Sgr~A$^*$, with an integrated polarization 
of the order of 1 per cent. The two main reflection nebulae are identified as the best targets for spatially-resolved X-ray 
polarization measurement, as they should produce $P~\gg$~10 per cent. The degree of polarization from Sgr~B2 and Sgr~C is a function
of the spatial location of these clouds and can constrain the distance of the reflecting nebulae to the central supermassive black 
hole. Measuring the polarization position angle of Sgr~B2 and Sgr~C with the GPD instrument at the focus of an X-ray optics 
would then pinpoint the illuminating source of the molecular clouds and determine if the tentative past flare from Sgr~A$^*$ is a 
viable source for the diffuse emission reported by \citet{Sunyaev1993}, \citet{Koyama1996} and \citet{Murakami2001}. The reconstruction 
of the light-curve of Sgr~A$^*$ outburst can be achieved if the detector has a field of view with sufficient enough resolution, 
such as the concept proposed previously for {\it NHXM}.

\section*{Acknowledgements}

We thank the anonymous referee for helpful comments. The authors also acknowledge support from the collaboration project between 
the Czech Science Foundation and Deutsche Forschungsgemeinschaft (GACR-DFG~13-00070J), as well as the COST-CZ LD12010 grant.


\bsp

\label{lastpage}

\end{document}